# Chemical reactivity imprint lithography on graphene: Controlling the substrate influence on electron transfer reactions

Qing Hua Wang[1], Zhong Jin[1], Ki Kang Kim[2], Andrew J. Hilmer[1], Geraldine L.C. Paulus[1], Chih-Jen Shih[1], Moon-Ho Ham[3], Javier Sanchez-Yamagishi[4], Kenji Watanabe[5], Takashi Taniguchi[5], Jing Kong[2], Pablo Jarillo-Herrero[4], and Michael Strano[1*]

[1]Department of Chemical Engineering, Massachusetts Institute of Technology, Cambridge, Massachusetts 02139, USA

[2]Department of Electrical Engineering and Computer Science, Massachusetts Institute of Technology, Cambridge, Massachusetts 02139, USA

[3]School of Materials Science and Engineering, Gwangju Institute of Science and Technology, Gwangju 500-712, South Korea

[4]Department of Physics, Massachusetts Institute of Technology, Cambridge, Massachusetts 02139, USA

[5]Advanced Materials Laboratory, National Institute for Materials Science, 1-1 Namiki, Tsukuba 305-0044, Japan

[*]Corresponding author. Email: strano@mit.edu

**Abstract**

The chemical functionalization of graphene enables control over electronic properties and sensor recognition sites. However, its study is confounded by an unusually strong influence of the underlying substrate. In this paper, we show a stark difference in the rate of electron transfer chemistry with aryl diazonium salts on monolayer graphene supported on a broad range of substrates. Reactions proceed rapidly when graphene is on $SiO_2$ and $Al_2O_3$ (sapphire), but negligibly on alkyl-terminated and hexagonal boron nitride (hBN) surfaces. The effect is contrary to expectations based on doping levels and can instead be described using a reactivity model accounting for substrate-induced electron-hole puddles in graphene. Raman spectroscopic mapping is used to characterize the effect of the substrates on graphene. Reactivity imprint lithography (RIL) is demonstrated as a technique for spatially patterning chemical groups on graphene by patterning the underlying substrate, and is applied to the covalent tethering of proteins on graphene.

**Introduction**

Graphene is a two-dimensional, atomically thin lattice of $sp^2$-bonded carbon atoms with exceptional electronic, mechanical, and thermal properties.[1,2] Modifying the basic electronic, chemical, and structural properties of graphene is important for incorporating graphene into a variety of applications including electronic devices, biosensors, and composite materials.[3] The chemical functionalization of graphene is important for enabling these applications, and has been explored via covalent[4,5] and noncovalent[6-8] schemes. The functionalization of graphene with aryl diazonium salts[4,9-16] results in the opening of a band gap[10,13,17-19] and shifting of the Fermi level,[10] which are both desirable for the fabrication of electronic devices. In addition, the functional groups on the diazonium moiety can be tailored by organic chemistry so that various chemical characteristics to be coupled to graphene.[9]

Graphene is strongly influenced by the underlying substrate. While SiO$_2$/Si substrates are compatible with device fabrication, their roughness and charged impurities lead to electron-hole charge fluctuations or puddles that scatter carriers and inhibit electronic device performance.[20,21] Graphene devices suspended over gaps exhibit the highest carrier mobilities,[22,23] but are not robust for practical use. Recently, single-crystal hexagonal boron nitride (hBN)[24,25] and self-assembled monolayers (SAMs) of hydrophobic molecules grafted on SiO$_2$ substrates[26-29] have been explored as alternate substrates for graphene electronics. Graphene on hBN, which is atomically smooth, chemically inert, and electrically insulating, has significantly lower electron-hole puddles and higher mobilities.[24,25] Graphene devices on SAMs also exhibit lower charge inhomogeneity and hysteresis[26,27] because the SAMs prevent dipolar contaminants from adsorbing on the substrate, prevent charge injection from the graphene to the dielectric interface, and screen the effect of charged impurities within the substrate.[26,27,29]

In this paper, we report that the substrate on which graphene rests strongly influences chemical reactions on its top surface. We can spatially control the chemical reactivity of graphene with micron-scale resolution to achieve wafer-scale patterning of chemical reactions on graphene. A previous report has shown differences in reaction for small mechanically exfoliated flakes of graphene on SiO$_2$ and hexamethyldisilazane (HMDS)-treated SiO$_2$.[30] In the current work, chemical vapour deposition (CVD)-grown graphene graphene is deposited on a variety of substrates and covalently functionalized with aryl diazonium salts. Using Raman spectroscopic mapping, we find that the substrate-induced electron-hole charge fluctuations in graphene greatly influence the chemical reactivity. Graphene on SiO$_2$ and Al$_2$O$_3$ (sapphire) substrates is highly reactive while graphene on an alkyl-terminated monolayer and hBN is much less reactive. We develop a new lithographic patterning technique, reactivity imprint lithography (RIL), where the underlying substrate is chemically patterned to achieve spatial control of graphene chemical reactivity. This method allows chemical reactions on graphene to be spatially patterned over large areas without the use of disruptive resist materials or etchants. We use RIL to spatially control the conjugation of enhanced green fluorescent protein (EGFP) on graphene directly from solution, demonstrating the advantages of the technique for producing structures for sensor and microarray applications.

**Results and discussion**

*Chemical reactivity of graphene on different substrates*

Monolayer graphene grown by chemical vapour deposition (CVD) on Cu foils[31,32] is transferred to several different substrates. Covalent functionalization via an electron transfer reaction with 4-nitrobenzenediazonium (4-NBD) tetrafluoroborate results in nitrobenzene groups covalently attached to the graphene lattice (Figure 1a). Figure 1b shows the substrates used in this work: 300 nm thick $SiO_2$ on Si wafer; self-assembled monolayer (SAM) of octadecyltrichlorosilane (OTS) on 300 nm $SiO_2$; mechanically exfoliated flake of 90 nm thick single crystal hexagonal boron nitride (hBN) deposited on 300 nm $SiO_2$; and a single crystal wafer of α-$Al_2O_3$ (sapphire, c-plane, (0001) orientation). The $SiO_2$ substrate was cleaned by oxygen plasma to generate a hydrophilic surface terminated with –OH groups.

Representative Raman spectra of graphene on each substrate before and after diazonium functionalization are shown in Figure 1b. The primary peaks are the G peak near 1580 cm$^{-1}$, the D peak near 1300-1350 cm$^{-1}$, and the 2D peak near 2600-2700 cm$^{-1}$.[33,34] The G and 2D peaks provide information about the level of doping, strain, and layer number,[33-36] and the D peak is activated by lattice defects,[37] including physical damage[36,38] and formation of sp$^3$ hybridization by covalent chemistry.[5,10] The integrated intensity ratio of the D and G peaks ($I_D/I_G$) is a measure of the concentration of covalent defect sites. In the spectra for pristine graphene in Figure 1b, which are normalized to the G peak height, the D peak is very small on all substrates, and differences are seen in the $I_{2D}/I_G$ ratios. After diazonium functionalization, prominent D peaks and small D´ peaks appear on $SiO_2$ and $Al_2O_3$ substrates, indicating significant formation of sp$^3$ bonds. On OTS and hBN substrates, very small D peaks appear, indicating sparse covalent functionalization. For all substrates, the G and 2D peaks are shifted up in position, and the 2D peak intensity is decreased.

The correlation of chemical reactivity with the hydrophobicity of the underlying substrate is shown in Figure 1c. In addition to the oxygen-plasma-cleaned bare $SiO_2$, we studied $SiO_2$ cleaned by piranha solution, which also produces a hydrophilic surface, and a sample that was used as-received. The hBN flakes were typically under 100 μm in size and were too small for macroscopic contact angle measurements. In general, the contact angle of the substrate appears to be inversely correlated with graphene chemical reactivity. Low contact angles indicate hydrophilicity due to polar chemical groups at the surface, which can induce electron-hole puddles in graphene, while high contact angles indicate nonpolar surfaces. Further analysis of Raman spectra is conducted to clarify the role of the substrate in changing the chemical reactivity of graphene.

*Analysis of Raman spectroscopic maps*

To account for spatial non-uniformity, two-dimensional Raman maps with 121 spectra each were taken in the same 10 μm x 10 μm sample areas before and after diazonium functionalization. The average peak parameters from fitting the peaks to Lorentzian functions are summarized in Table 1. Histograms of the $I_D/I_G$ ratio in Figure 2a show very low initial defect concentrations. After diazonium functionalization, the centres of the distributions have increased to ~1.2 for $Al_2O_3$ and ~1.4 for $SiO_2$, indicating a relatively high degree of covalent functionalization. The histograms are also wider, suggesting an increased spatial inhomogeneity. For hBN and OTS, the $I_D/I_G$ ratio has only slightly increased to ~0.25, indicating much lower reactivity.

Scatter plots of the Raman peak parameters are shown in Figures 2b-e. Data from mechanically exfoliated monolayer graphene doped by electrostatic gating are included on these plots as comparisons.[35,39] In Figure 2b, the full width at half maximum (FWHM) of the G peak ($\Gamma_G$) is plotted against the position of the G peak ($\omega_G$). The dashed trend line indicates increasing n- or p-doping leads to narrowing of the G peak and increase of the G peak position.[34,39,40] This trend line has been shifted upward to accommodate the wider G peak in CVD graphene. Pristine graphene on each of the substrates generally follows the doping trend line, with hBN closer to the undoped region and $Al_2O_3$ closer to the more doped region. However, electron and hole doping and cannot be distinguished from this plot, and also graphene that is uniformly electron- or hole-doped cannot be distinguished from graphene with many electron- and hole-doped charge puddles. After diazonium functionalization, $\omega_G$ is upshifted for all substrates, suggesting increased doping, while $\Gamma_G$ is also much higher for $SiO_2$ and $Al_2O_3$, suggesting increased disorder.[41]

The G and 2D peak positions ($\omega_G$ and $\omega_{2D}$) are plotted against each other in Figure 2c along with comparison data[39] to distinguish between n- and p-doping trends. The unfunctionalized graphene in our samples lies in the slightly p-doped region of this plot, with the hBN surface being less doped. However, graphene on $Al_2O_3$ is on the p-doping branch. Graphene on all the substrates after functionalization is further along the p-doping branch. However, covalent defects are expected to cause deviations from these Raman trends due to doping for pristine graphene. The p-doping after reaction has contributions from the covalent bond formation itself and from the non-covalent adsorption of the diazonium cation and oligomers.[12,13,18]

The FWHM of the 2D peak ($\Gamma_{2D}$) is plotted against its position ($\omega_{2D}$) in Figure 2d. Since the 2D peak position shifts in opposite directions for electron and hole doping (Figure 2c), the presence of electron-hole puddles significantly smaller than the Raman laser spot size would result in a broadened 2D peak. In our Raman system, the laser spot size is ~0.9 μm in diameter, and the sizes of electron-hole puddles have been measured to be about 5-10 nm in diameter for graphene on $SiO_2$ and ~100 nm for graphene on hBN.[25] Therefore we propose that a higher $\Gamma_{2D}$ is correlated with higher amplitudes of charge fluctuations. Graphene on $SiO_2$ exhibits the highest $\Gamma_{2D}$ values, while graphene on hBN has the lowest. This trend is in general agreement with the amplitudes of charge fluctuations on $SiO_2$ and hBN measured by scanning tunneling spectroscopy.[25] On OTS, the $\Gamma_{2D}$ is a bit higher than on hBN and notably lower than on $SiO_2$.

The integrated area intensity ratio $I_{2D}/I_G$ is plotted against $\omega_G$ in Figure 2e, with additional comparison data for gated pristine graphene adapted from Ref.[35] showing that the $I_{2D}/I_G$ ratio decreases while the $\omega_G$ increases for increasing n- and p-doping. Graphene on hBN is closest to the undoped region of the plot, followed by OTS, SiO$_2$, and finally Al$_2$O$_3$ at the more highly doped region. (Although the peak intensities on Al$_2$O$_3$ have not been corrected for optical interference effects from the different substrate,[42] the peak positions are accurate.) After diazonium functionalization, the data points from all substrates move further along the doping trend line. Again, we observe that the diazonium functionalization increases the p-doping of the graphene.

Graphene on the various substrates displays different extents of overall p-doping and apparent intensities of electron-hole charge fluctuations. Graphene on hBN is the least doped with the lowest degree of charge fluctuations, followed by OTS. In contrast, graphene on SiO$_2$ and Al$_2$O$_3$ are more highly p-doped, and on SiO$_2$ the $\Gamma_{2D}$ is the highest, indicating the most broadening of the 2D peak from electron-hole puddles. After reaction, graphene on all substrates showed increased p-doping. For the substrates with a low degree of sp$^3$ hybridization, the p-doping arises from diazonium molecules noncovalently deposited on graphene. The role of electron-hole puddles in the reactivity of graphene is discussed below.

*Spatial patterning of chemical reactivity*

In reactivity imprint lithography (RIL), a substrate with OTS micropatterned[43,44] on SiO$_2$ is used to spatially control the chemical reactivity of graphene as illustrated in Figure 3a. The patterned surface in the topographic atomic force microscopy (AFM) image of Figure 3b has ~2 μm OTS lines and ~7 μm SiO$_2$ gaps. Graphene is then transferred onto this substrate and functionalized by diazonium salts. Figure 3c shows the resulting spatial Raman map of $I_D/I_G$. The narrower regions of low functionalization correspond to graphene over OTS-covered areas and the wider stripes of high functionalization the SiO$_2$ regions.

The $I_D/I_G$ spatial profile at the edge of a stripe was fit using an integral Gaussian distribution in Figure 3d (see details in Supplemental Information). The variance of this fit reflects the sharpness of the transition between the on-OTS and on-SiO$_2$ regions, and is about 0.85 μm. The $I_D/I_G$ profile for graphene across the edge of a flake of hBN is plotted and fitted similarly in Figure 3e, with a variance of 0.76 μm. These variances are comparable to the 0.71 μm diagonal of the pixel size (0.5 μm x 0.5 μm) and the ~0.9 µm laser spot size. Therefore the measured resolution of the RIL patterns is limited by the optical characterization technique, and the true resolution of the chemical patterning is primarily determined by the spatial resolution of the substrate patterning technique and spatial size of electron-hole puddles on a given substrate, which the data indicate as less than 1 µm.

*Patterned attachment of proteins on graphene*

Spatial control of surface chemistry is important for biological applications such as microarrays, biosensors, and tissue engineering. Many important macromolecules such as proteins, antibodies or DNA are not compatible with conventional lithographic techniques. Reactivity imprint lithography allows these biomolecules to be attached to graphene as the final processing step in aqueous solution. The patterning of biomolecules on graphene using RIL is schematically illustrated in Figure 4a. CVD graphene is transferred to an OTS-patterned substrate and functionalized by 4-carboxybenzenediazonium tetrafluoroborate. Then, the graphene is reacted with $N_\alpha,N_\alpha$-bis(carboxymethyl)-L-lysine hydrate (NTA-$NH_2$) followed by reaction with $NiCl_2$ to complex the $Ni^{2+}$ ions with the NTA structure. Finally, the sample is incubated with a solution of polyhistidine (His)-tagged enhanced green fluorescent protein (EGFP) to form the graphene-NTA-Ni-His-EGFP complex.

Attachment of the carboxybenzene group is shown by attenuated total reflectance infrared (ATR-IR) spectra of the pristine CVD graphene (blue curve) and functionalized graphene (red curve) in Figure 4b. Vibrations from carboxyl groups are seen at ~1730 cm$^{-1}$ (C=O stretching) and ~3330 cm$^{-1}$ (O–H stretching). Confocal fluorescence microscopy after incubation in EGFP shows bright green stripes confirming the spatial patterning of the protein tethering reaction (Figure 4c). The wider, bright lines correspond to graphene resting on $SiO_2$ where the higher concentration of diazonium attachment sites results in a high coverage of EGFP. The narrower, dark lines correspond to graphene resting on OTS where the low reactivity results in fewer EGFP. The inset shows the fluorescence intensity profile along the white line. This tethering scheme is very robust due to the covalent attachment site, and is also chemically reversible due to the metal ion chelation, compared to a previous report of proteins patterned on graphene by physisorption.[45]

*Reactivity model – the influence of electron puddles*

To explain the chemical reactivity of graphene on the different substrates, we use a model describing the reaction kinetics from electron transfer theory as a function of the Fermi level of graphene and relating the reacted site density to an experimentally measurable Raman $I_D/I_G$ ratio. Due to the overlap between graphene and diazonium states, the electron transfer theory below shows that the reactivity increases for increasingly n-doped graphene and is negligible for p-doped graphene. In Figure 5a, a schematic illustration shows how a graphene sheet that is p-doping overall with a high electron-hole charge fluctuation amplitude can have much higher reactivity due to the locally n-doped puddles.

In a first order, electron transfer reaction model, the density of reacted lattice sites σ is:

$$\sigma = \rho_C \left(1 - e^{-\left(\frac{k_{ET}[D]_S}{\rho_C}\right)t}\right) \tag{1}$$

where $\rho_C$ is the number of carbon atoms per unit area in graphene, $[D]_S$ is the concentration of diazonium ions, and $t$ is the reaction time. The reaction rate is limited by the electron transfer rate from graphene to diazonium, as is the case for carbon nanotubes.[46] The rate constant $k_{ET}$ is described using Gerischer-Marcus theory:[47]

$$k_{ET} = \nu_n \int_{E_{F,R}}^{E_{F,G}} \varepsilon_{ox}(E) DOS_G(E) W_{ox}(E) dE \tag{2}$$

where $E_{F,G}$ and $E_{F,R}$ are the Fermi levels of graphene and the redox species, respectively, and $DOS_G(E)$ is the electronic density of states of graphene. The electron transfer frequency $\nu_n$ and integral prefactor $\varepsilon_{ox}$ are treated as a single fitting parameter $\nu_n\varepsilon_{ox}$. The distribution of oxidized states of the solvated diazonium molecule $W_{ox}(E)$ is:

$$W_{ox}(E) = \frac{1}{\sqrt{4\pi\lambda kT}} \exp\left(-\frac{(E-(E_{redox}+\lambda))^2}{4\lambda kT}\right) \tag{3}$$

where $E_{redox} = -0.39$ eV is the redox potential of the diazonium molecule relative to the intrinsic Fermi level of graphene (–4.66 eV), assumed to be a similar value as that of 4-chlorobenzenediazonium salt.[46] The reorganization energy $\lambda$ is ~0.7 eV for single-walled carbon nanotubes,[46] and is assumed to be similar for graphene.

The density of reacted sites $\sigma$ can be related to the $I_D/I_G$ ratio by Lucchese et al:[38]

$$\frac{I_D}{I_G} = C_A \frac{r_A^2 - r_S^2}{r_A^2 - 2r_S^2} \left[ \exp\left(-\frac{\pi r_S^2}{L_D^2}\right) - \exp\left(-\frac{\pi(r_A^2 - r_S^2)}{L_D^2}\right) \right] + C_S \left[ 1 - \exp\left(-\frac{\pi r_S^2}{L_D^2}\right) \right] \quad (4)$$

where the distance between defects is $L_D = 1/\sqrt{\sigma}$. Around each defect site is a structurally damaged region with radius $r = r_S$ and a region within $r = r_S$ and $r = r_A$ that is primarily responsible for an increase in the D peak. In Ref. [38], changes in $I_D/I_G$ are caused by ion bombardment damage, but covalent functionalization with diazonium salts results in a slightly different behaviour of $I_D/I_G$,[10] so we used smaller values of $r_S = 0.07$ nm and $r_A = 1.0$ nm because a covalent attachment site is much less disruptive to the lattice than an ion bombardment defect. The parameters $C_A$ and $C_S$ were similar to values used in Ref. [38]. Combining equations (1)–(4) results in a curve showing the $I_D/I_G$ after diazonium functionalization as a function of graphene $E_F$ with $v_n \varepsilon_{ox}$ as the fitting parameter.

The model curve is plotted alongside experimental data from several samples of graphene on different substrate materials in Figures 5b-c. To obtain the average Fermi level, the Raman $I_G/I_{2D}$ ratio was used:[48]

$$\sqrt{\frac{I_G}{I_{2D}}} = C(\gamma_{e\text{-ph}} + 0.07|E_{F,\text{avg}}|) \quad (5)$$

where $\gamma_{e\text{-ph}} = $ ~33 meV is the average energy of electron scattering due to phonon emission and $C = $ ~10 eV$^{-1}$.[48] We have used the $\omega_{2D}$ vs. $\omega_G$ data (Figure 2c) to determine $E_{F,\text{avg}} < 0$. However, the hole-doped data show little agreement with the model in Figure 5b. To account for electron-hole puddles as illustrated in Figure 5a, we note that the reactivity is instead dominated by the sum of the Fermi level $E_F$ and the amplitude of the puddle, which should be proportional to the increase in $\Gamma_{2D}$ compared to the case with negligible puddle influence. Specifically, the effective Fermi level of the n-doped puddles is:

$$E_{F,n} = E_{F,\text{avg}} + \alpha(\Gamma_{2D} - \Gamma_{2D,0}) \quad (6)$$

where $\alpha$ is a proportionality constant, and $\Gamma_{2D,0}$ is the FWHM of the 2D peak for graphene with no charge puddles. In Figure 5c, the data points were shifted using $\alpha$ = 0.08 eV cm and $\Gamma_{2D,0}$ = 26 cm$^{-1}$, and the model curve is plotted with $v_n\varepsilon_{ox}$ = 0.105. (Note that $\alpha < 0$ for $E_{F,p}$ in the p-doped puddles.) After the adjustment in equation (6), the data are better described by the model.

Considering the Raman spectral analysis and the modeling results above, we can summarize the effects of the different substrates on graphene's chemical reactivity. Graphene on hBN and OTS has low electron-hole fluctuations and hence lower diazonium reactivity, while graphene on hydrophilic SiO$_2$ (plasma-cleaned and piranha-cleaned) and Al$_2$O$_3$ has higher charge fluctuations that result in more n-doped reactive regions. The charge fluctuations on SiO$_2$ are caused by charged impurities in the substrate and polar adsorbates on the surface, so adding the OTS monolayer decreases the fluctuations by increasing the distance between graphene and the charged impurities and by reducing the adsorption of polar adsorbates such as water. An unknown film of organic contamination likely covers as-received SiO$_2$ substrates and serves a similar role as the OTS monolayer. Although the Al$_2$O$_3$ substrates are single crystals in the bulk, their surfaces are likely to be similar to the amorphous SiO$_2$ substrates.

The Fermi level offsets calculated in equation (6) are larger than the magnitude of electron-hole fluctuations reported earlier for mechanically exfoliated single-crystal graphene.[25] This difference may be explained by grain boundaries and other contaminants in the CVD graphene that can increase the reactivity for a lower Fermi level shift. Furthermore, the 2D Raman peaks from graphene with high electron-hole fluctuations would also have lower intensities, causing the $E_{F,\text{avg}}$ calculated from equation (5) to be further from neutrality and requiring a larger shift in equation (6) to fit the $I_D/I_G$. The hydrophobicity of the substrate is an initial predictor of the chemical reactivity as shown in Figure 1c because the surface energy of the substrate relates to the presence of charged impurities and polar surface groups that can induce electron-hole charge fluctuations in graphene.

**Conclusions**

In summary, the effect of the underlying substrate on the chemical reactivity of graphene has been explored. Graphene on $SiO_2$ and $Al_2O_3$ is more reactive toward covalent functionalization by aryl diazonium salts than graphene on hBN and on an alkyl-terminated monolayer. The reactivity contrast is attributed to higher amplitudes of substrate-induced electron-hole charge fluctuations for graphene on $SiO_2$ and $Al_2O_3$. Micron-scale spatial control of graphene's chemical reactivity was demonstrated by chemically patterning the substrate prior to deposition of graphene. Due to the versatility and chemical tailorability of reactivity imprint lithography, the technique can be expanded in many directions for the modification and manipulation of graphene. This chemical patterning technique was also applied to the spatial patterning of protein molecules on graphene, demonstrating the potential for applications in biosensing.

## Methods

**Graphene synthesis and transfer.** Copper foil substrates (25 μm, 99.8%, Alfa Aesar) were first annealed (1000°C, 30 min, 10 sscm hydrogen, ~330 mtorr total pressure) followed by graphene synthesis (1000°C, 40 min, 15 sccm methane and 50 sccm hydrogen, ~1.5 torr total pressure). Graphene on Cu was covered in poly(methyl methacrylate) (950PMMA, A4, MicroChem) by spin coating (3000 rpm, 1 min) and dried in air (30 min). Graphene on the reverse side was removed by reactive ion etching (Plasmatherm RIE, 100 W, 7 mtorr oxygen, 5 min). The PMMA-graphene-Cu stack was placed on the surface of Cu etchant (6 M HCl and 1 M $CuCl_2$ in water). After Cu etching (~30 min), the PMMA-graphene layer was scooped out with a clean wafer and floated onto several baths of ultrapure water for rinsing. It was scooped out with the target substrate and dried in air overnight before immersion in several baths of clean acetone to dissolve the PMMA, followed by rinsing in isopropanol and drying with ultrapure nitrogen.

**Surface preparation of wafer substrates.** Silicon wafers with 300 nm $SiO_2$ were ultrasonically cleaned in sequential baths of acetone, isopropanol, and water, followed by additional surface treatments. (1) Plasma-cleaned samples: exposed to oxygen plasma (AutoGlow Plasma System, Glow Research) for 10-30 min at 200 W power and 0.5 torr. (2) Piranha-cleaned samples: immersed in a piranha solution (3:1 solution of sulfuric acid to 30% hydrogen peroxide) for 15 minutes and rinsed in ultrapure water. (Warning: piranha solution is a strong oxidizing agent and should be handled with extreme care!) (3) As-received samples: no additional treatment. Sapphire wafers ($\alpha$-$Al_2O_3$, c-plane, 0.5 mm thick, MTI Corp.) were ultrasonically cleaned in acetone and isopropanol.

**OTS monolayer on $SiO_2$.** Octadecyltrichlorosilane (OTS) (Sigma-Aldrich, 90+%) self-assembled monolayers were formed on freshly plasma-cleaned $SiO_2$ substrates in OTS solution (10 mM in toluene) overnight in a closed vial, then rinsed in fresh toluene and blown dry with nitrogen.

**Surface patterning of substrates.** OTS patterns were formed on freshly plasma-cleaned $SiO_2$ substrates by printing with polydimethylsiloxane (PDMS) stamps. Master patterns were formed by electron beam lithography in PMMA resists on silicon wafers. PDMS (10:1 mass ratio of base to curing agent, Dow Corning Sylgard 184) was poured into the master patterns, degassed in vacuum for 45 min, and cured at 100°C for 2 h on a hotplate. The stamps were inked by spin coating 10 mM OTS in anhydrous toluene (3000 rpm, 30 s). The stamps were gently brought into contact with the substrates for 60 s.

**hBN preparation.** The hBN flakes used in this study were prepared by mechanical exfoliation of an ultrapure single crystal of hBN on piranha-cleaned $SiO_2$/Si substrates. The hBN crystal was grown by the method described previously.[49]

**Diazonium functionalization of graphene.** Graphene samples were immersed in aqueous solution of 10 mM 4-nitrobenzenediazonium tetrafluoroborate (4-NBD) and 0.5 wt% sodium dodecyl sulfate (SDS) with constant stirring at ~35°C. Most samples were reacted for 16.5 hr to reach full reaction conversion, while the sample in Figure 4 was reacted for 1.5 hr to improve $I_D/I_G$ spatial contrast. After reaction, the samples were rinsed in ultrapure water and blown dry with nitrogen.

**Raman spectroscopy and mapping.** Raman spectroscopy was performed on a Horiba Jobin Yvon LabRAM HR800 system using a 633 nm excitation laser, 100X objective lens with ~1 μm diameter spot size, and motorized XYZ stage. The G, 2D, and D peaks were fit to Lorentzian functions.

**Contact angle.** The contact angles of the substrates were measured using a Ramé-Hart goniometer and 2 μL sessile droplets of ultrapure water. Several droplets were measured in different sample locations and the results were averaged.

**Atomic force microscopy.** AFM imaging was conducted on an Asylum Research MFP-3D system in AC (noncontact) mode using silicon probes (Olympus OMCL-AC240TS). Images were processed using the Gwyddion software package.

**Binding of proteins on graphene.** Graphene samples on OTS-patterned $SiO_2$ substrates were immersed in an aqueous solution of 1 wt% SDS and 50 μM 4-carboxybenzenediazonium tetrafluoroborate and stirred at 45°C for 12 h. They were then immersed in a phosphate buffered solution (pH 8.3) with 100 μM of $N_α,N_α$-bis(carboxymethyl)-L-lysine hydrate (NTA-$NH_2$) at room temperature for 8 h, followed by an aqueous solution of 20 μM $NiCl_2$ at room temperature for 4 h to complex the $Ni^{2+}$ ions to the NTA structure. Then they were immersed in an aqueous solution of 1 μM polyhistidine (His)-tagged enhanced green fluorescent protein (EGFP) at room temperature for 1 h. Between each step described above, the sample was rinsed with water, acetone, and isopropanol and blown dry with ultrapure nitrogen. Attenuated total reflectance infrared (ATR-IR) spectra were obtained using a Thermo Nicolet 4700 Spectrometer. Confocal fluorescence microscopy images were captured using a Zeiss LSM 710 NLO with 633 nm laser excitation.

**Acknowledgements**

This work was primarily funded by 2009 US Office of Naval Research Multi University Research Initiative (MURI) grant on Graphene Advanced Terahertz Engineering (GATE) at MIT, Harvard and Boston University. M.S. Strano appreciates characterization support from the Institute of Soldier Nanotechnologies at MIT funded by a grant from the US Army Research Office. J. Sanchez-Yamagishi and P. Jarillo-Herrero acknowledge support by NSF CAREER Award DMR-0845287. K.K. Kim acknowledges the NSF award number DMR-0845358 and the Materials, Structures and Device (MSD) Center of the Focus Center Research Program (FCRP) at the Semiconductor Research Corporation. The authors thank M.K. Mondol of the MIT Scanning Electron Beam Lithography facility for assistance.

**Author contributions**

Q.H.W. designed and conducted the substrate and patterning experiments, performed Raman spectroscopy, AFM, and data analysis. Z.J. performed the protein attachment, ATR-IR, and fluorescence imaging. A.J.H., Q.H.W. and M.S.S. devised the model. K.K.K. synthesized the CVD graphene. K.W. and T.T. synthesized the hBN crystal. J.S.-Y. exfoliated the hBN crystal. G.L.C.P., C.-J.S. and M.-H.H. conducted additional experiments. Q.H.W. and M.S.S. wrote the manuscript. All authors contributed to discussion and interpretation of results.

# Figures

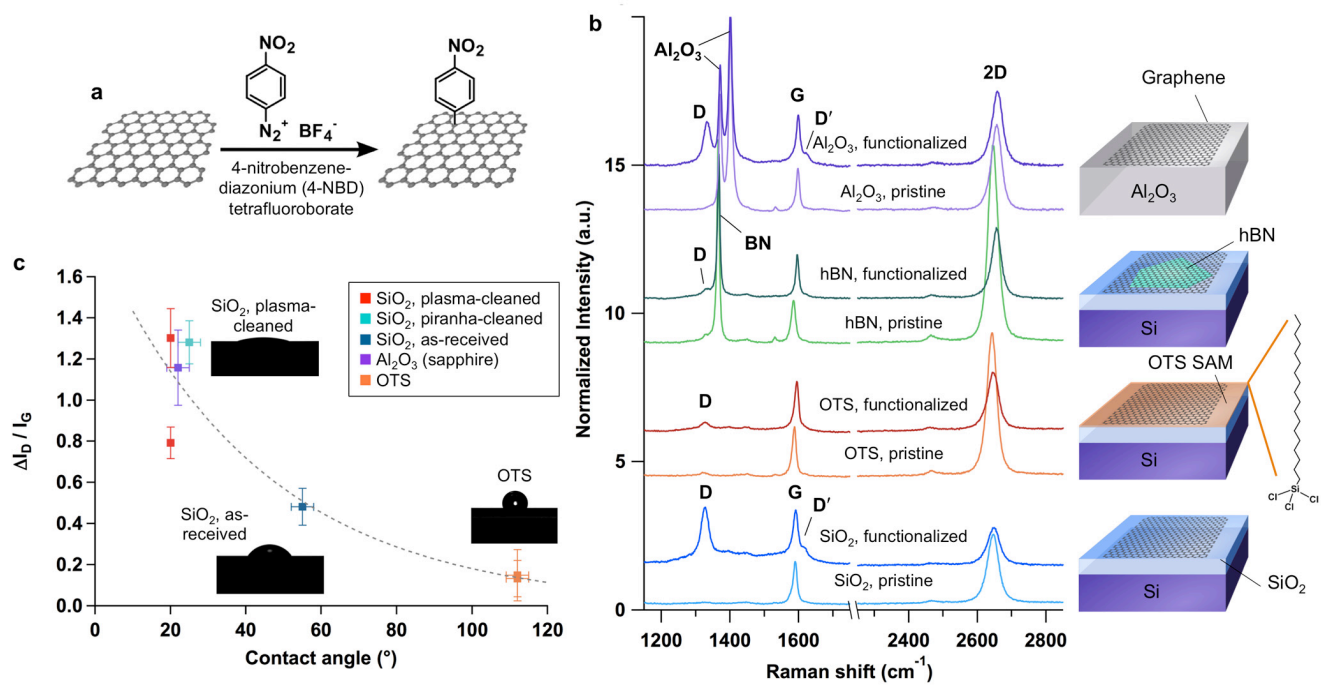

**Figure 1 | Chemical reactivity of graphene supported on different substrates. a,** Reaction scheme of covalent chemical functionalization of graphene by 4-nitrobenzenediazonium (4-NBD) tetrafluoroborate. **b,** Representative Raman spectra of chemical vapour deposition (CVD)-grown graphene deposited on different substrate materials before and after diazonium functionalization, normalized to the G peak height. These substrates are, from bottom to top: 300 nm thick $SiO_2$ on Si; $SiO_2$ functionalized by an octadecyltrichlorosilane (OTS) self-assembled monolayer (SAM); single crystal hexagonal boron nitride (hBN) flakes deposited on $SiO_2$; and single crystal α-$Al_2O_3$ (c-face sapphire). **c,** Change in $I_D/I_G$ Raman intensity ratio after diazonium functionalization (difference between functionalized and unfunctionalized ratios) plotted as a function of the water contact angle of the substrate prior to graphene deposition.. The dashed line is an exponential fit of the data. Raman spectra were taken at 633 nm laser excitation wavelength.

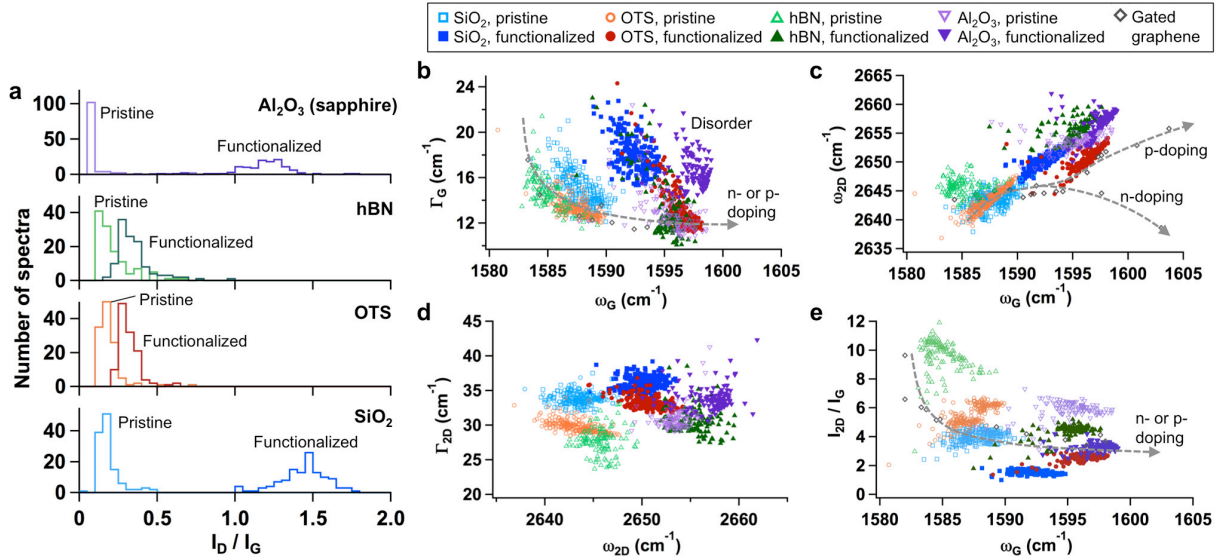

**Figure 2 | Raman spectroscopy peak parameter analysis.** Spatial Raman maps were measured for graphene supported on each substrate, in the same 10 μm x 10 μm regions before and after diazonium functionalization, with 121 spectra in each map. **a,** Histograms of $I_D/I_G$ ratios before and after functionalization. A low degree of covalent functionalization (small increase in $I_D/I_G$) is seen on OTS and hBN, and a much higher degree on $SiO_2$ (plasma-cleaned) and $Al_2O_3$. **b-e,** Scatter plots of Raman peak parameters with data points adapted from pristine, mechanically exfoliated graphene doped by electrostatic gating and dashed lines added to guide the eye included as comparison.[35,39] **b,** G peak full width at half maximum (FWHM, $\Gamma_G$) vs. G peak position ($\omega_G$). The comparison data from Ref.[39] is shifted up to fit the higher FWHM of CVD graphene. Before reaction, graphene follows the doping trend, but highly functionalized samples significantly deviate above the curve. **c,** 2D peak position ($\omega_{2D}$) vs. G peak position ($\omega_G$), with additional data points adapted from Ref.[39] for distinguishing n-doped and p-doped exfoliated monolayer graphene, shifted to account for $\omega_{2D}$'s dependence on excitation laser wavelength. Diazonium-functionalized graphene in our experimental data is p-doped, but deviates left from the trend of pristine, gated graphene. **d,** 2D peak FWHM ($\Gamma_{2D}$) vs. 2D peak position ($\omega_{2D}$), showing clearly distinguished clusters for each substrate before and after functionalization. Increasing $\Gamma_{2D}$ values before functionalization reflect inhomogeneous broadening due to electron-hole charge fluctuations. **e,** $I_{2D}/I_G$ intensity ratio vs. G peak position ($\omega_G$), with comparison data adapted from Ref.[35] showing the doping trend. Raman spectra were taken at 633 nm laser excitation wavelength.

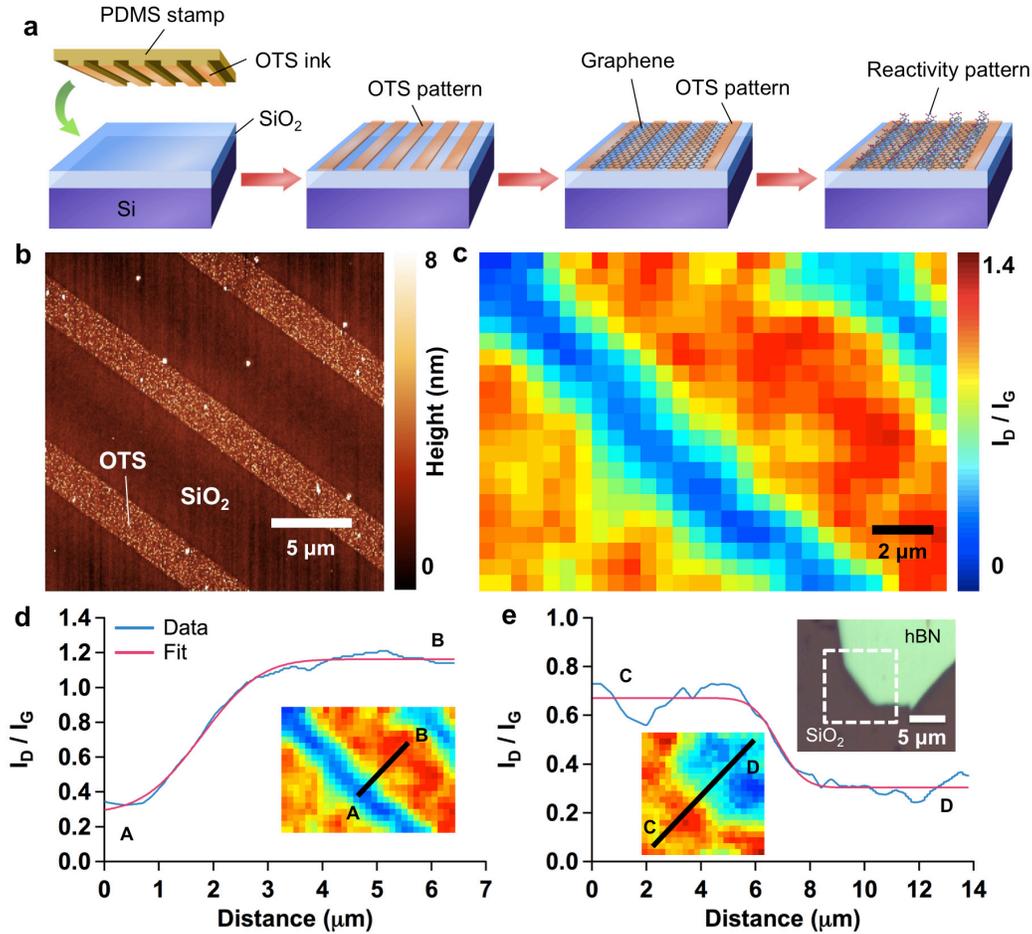

**Figure 3 | Spatial control of reactivity of graphene on patterned substrates. a**, Schematic illustration of reactivity imprint lithography (RIL). The SiO$_2$ substrate is patterned by a polydimethylsiloxane (PDMS) stamp inked with octadecyltrichlorosilane (OTS). Graphene is transferred over the OTS-patterned substrate and reacted with 4-nitrobenzenediazonium tetrafluoroborate. **b**, Atomic force microscopy (AFM) topographic image of the OTS stripes (narrower raised regions) on SiO$_2$ before graphene deposition. **c**, Raman spatial map of $I_D/I_G$ intensity ratio after diazonium functionalization. The narrow, mildly functionalized stripes correspond to the regions over the OTS pattern and the wide, strongly functionalized stripes correspond to the regions over the SiO$_2$ gaps. **d**, Spatial profile of the $I_D/I_G$ for the stripe pattern (blue curve) along the line 'A–B' in the Raman map (inset), and a fit to an integrated Gaussian function with a variance of 0.85 μm. **e,** A spatial Raman map (lower left inset) was measured for a region of graphene covering both SiO$_2$ and a flake of hBN (white box in optical image in upper right inset). The $I_D/I_G$ spatial profile along the line 'C–D' is shown along with the integrated Gaussian fit, which has a variance of 0.76 μm.

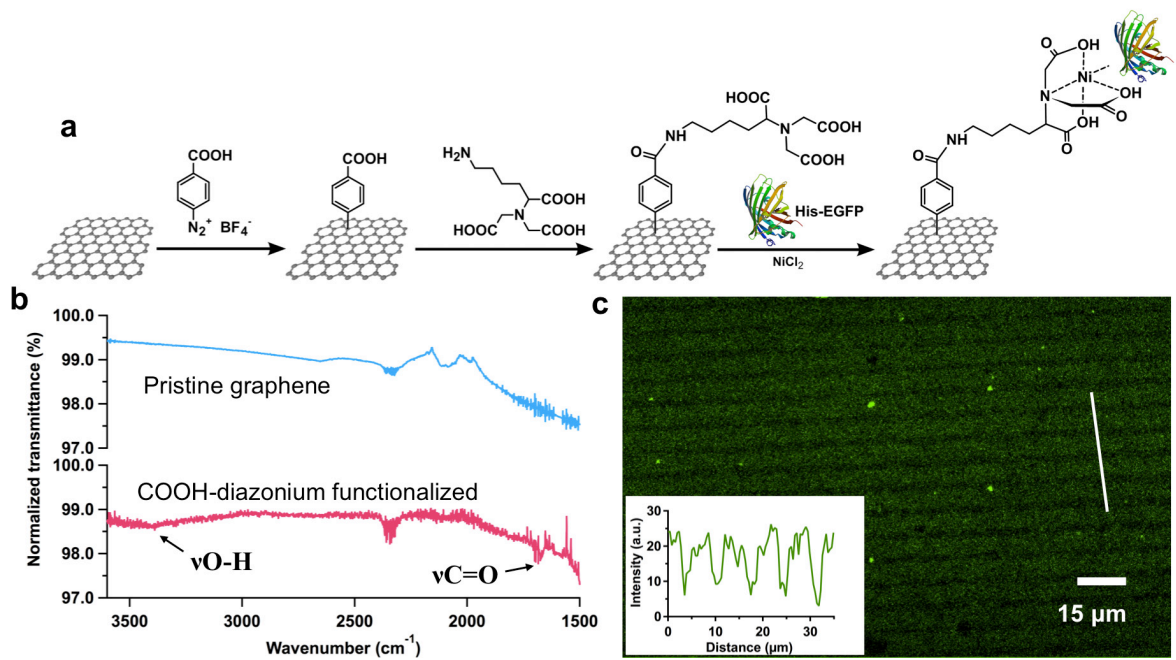

**Figure 4 | Patterning of proteins on graphene. a,** Schematic illustration of the protein attachment chemistry. The graphene is covalently functionalized with 4-carboxybenzenediazonium tetrafluoroborate, then Nα,Nα-bis(carboxymethyl)-L-lysine hydrate (NTA-NH$_2$). Reaction with NiCl$_2$ causes Ni$^{2+}$ ions to complex to the covalently attached structures, and link to polyhistidine (His)-tagged enhanced green fluorescent protein (EGFP). Image of EGFP from the RCSB PDB (www.pdb.org) from Ref. [50]. **b,** Attenuated total reflectance infrared (ATR-IR) spectra of pristine CVD graphene (blue curve) and COOH-diazonium functionalized CVD graphene (red curve), showing O–H and C=O vibrations from carboxyl groups. **c,** Confocal fluorescence microscope image of enhanced green fluorescent protein (EGFP) attached to graphene resting on a substrate with alternating stripes of bare SiO$_2$ and OTS patterned on graphene. The bright green stripes, indicating a higher concentration of EGFP attachment, corresponds to graphene resting on bare SiO$_2$, while the darker stripes correspond to graphene resting on OTS-patterned regions where very little EGFP was able to attach. Inset: Intensity profile of fluorescence along the white line indicated in (c).

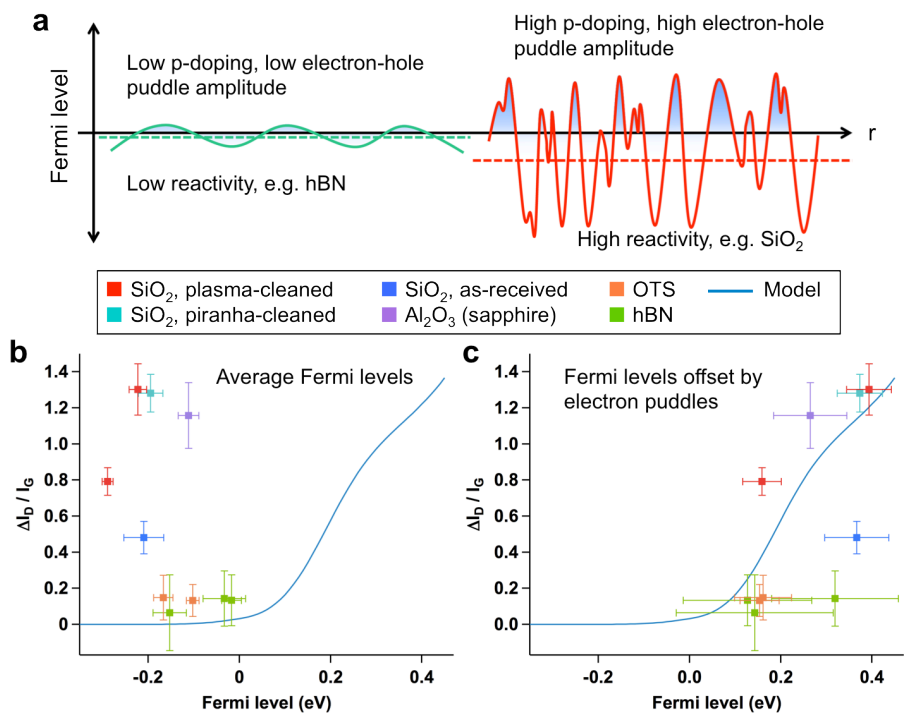

**Figure 5 | Modeling of substrate-influenced reactivity. a,** Schematic illustration of the role of electron-hole charge fluctuations in graphene reactivity. The solid curves indicate the spatial variation of local Fermi level in charge puddles, while the solid lines indicate the average Fermi level. The green curve (left) represents graphene on a substrate that causes it to be mildly p-doped with small charge flunctuations, and the red curve (right) represents higher p-doping and large charge fluctuations. According to electron transfer theory, the n-doped puddles have higher reactivity toward diazonium functionalization and the p-doped puddles have very low reactivity. **b,** Experimental data from graphene on various substrates are plotted alongside the curve from the electron transfer model for the initial graphene Fermi level ($E_F$) and change in $I_D/I_G$ ratio after diazonium functionalization. The experimental average $E_F$ values are calculated from the $I_{2D}/I_G$ ratios before functionalization.[35] Each experimental point is the average value on a particular sample taken from 121 Raman spectra in a map, and the error bars represent the standard deviation. The average doping for all samples is p-type, and does not agree with the model. **c,** The average $E_F$ values are offset by considering the FWHM of 2D peaks, which reflects inhomogeneous broadening due to electron-hole charge fluctuations, to reflect the maximum n-doping.

|  | $\omega_G$ (cm$^{-1}$) | $\Gamma_G$ (cm$^{-1}$) | $\omega_{2D}$ (cm$^{-1}$) | $\Gamma_{2D}$ (cm$^{-1}$) | $I_D / I_G$ | $I_{2D} / I_G$ |
|---|---|---|---|---|---|---|
| **SiO$_2$, pristine** | 1588.6 | 14.4 | 2644.1 | 33.7 | 0.11 | 4.24 |
| **SiO$_2$, functionalized** | 1591.9 | 18.1 | 2649.8 | 36.1 | 1.42 | 1.64 |
| **OTS, pristine** | 1588.3 | 12.7 | 2644.8 | 29.2 | 0.12 | 6.20 |
| **OTS, functionalized** | 1596.7 | 12.4 | 2651.1 | 33.0 | 0.25 | 2.66 |
| **hBN, pristine** | 1584.7 | 14.5 | 2645.6 | 27.8 | 0.13 | 9.88 |
| **hBN, functionalized** | 1595.6 | 12.1 | 2655.8 | 30.4 | 0.27 | 4.51 |
| **Al$_2$O$_3$ (sapphire), pristine** | 1595.6 | 12.5 | 2653.7 | 30.7 | ~0 | 6.01 |
| **Al$_2$O$_3$ (sapphire), functionalized** | 1598.0 | 16.3 | 2657.6 | 33.6 | 1.16 | 3.31 |

**Table 1 | Summary of graphene Raman peak parameters before and after diazonium functionalization.** The average values for key Raman peak parameters are summarized for pristine and functionalized graphene on SiO$_2$ (plasma-cleaned), OTS, hBN, and Al$_2$O$_3$ (sapphire) substrates. The parameters shown are the peak positions of G and 2D peaks ($\omega_G$ and $\omega_{2D}$) and full widths at half maximum (FWHM) of G and 2D peaks ($\Gamma_G$ and $\Gamma_{2D}$) and 2D/G and D/G integrated intensity ratios ($I_D/I_G$ and $I_{2D}/I_G$).

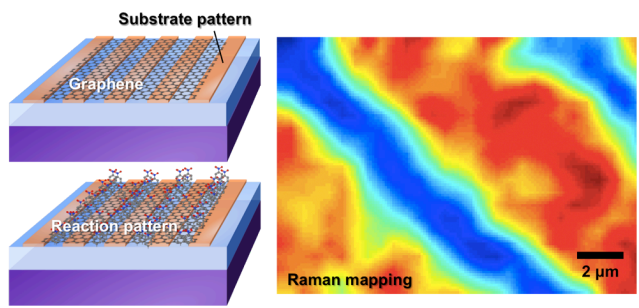

**Graphical Abstract Image**

**Caption:** Schematic illustration of graphene resting on a patterned substrate, which results in a chemical reaction pattern in graphene that can be imaged by Raman mapping.